\documentclass[11pt,a4paper]{article}
\usepackage[english]{babel}
\usepackage[margin=1in]{geometry}
\usepackage{multicol}
\usepackage{background}
\usepackage[math]{blindtext}
 
 \usepackage{amsmath}
 \usepackage[cmintegrals]{newtxmath}

 \usepackage{fonttable}
 \usepackage{url}
 \usepackage{multicol}
 \usepackage{booktabs}
 \usepackage{lipsum}
 \usepackage{colortbl}
 \usepackage{multicol}
 \usepackage{multirow}
 \usepackage{kantlipsum} 
 \usepackage{mwe} 
 \definecolor{mycolor}{rgb}{0.92,0.95,0.86}
 \definecolor{mycolor1}{RGB}{255, 204, 150}
 \definecolor{mycolor2}{RGB}{97, 48, 06}

\usepackage{amsmath}
\usepackage[cmintegrals]{newtxmath}
\usepackage{url}
\usepackage{multicol}
\usepackage{booktabs}
\usepackage{lipsum}
\usepackage{colortbl}
\usepackage{stfloats}
\usepackage{multicol}
\usepackage{multirow}
\usepackage{kantlipsum} 
\usepackage{mwe} 
\usepackage{float}
\usepackage{subfig}

\usepackage{xcolor}
\usepackage{caption}

\usepackage{xr}

\usepackage[ruled,linesnumbered]{algorithm2e}

\SetBgScale{4}
\SetBgColor{gray}
\SetBgAngle{90}
\SetBgContents{arXiv: some other text goes here}
\SetBgPosition{-1.5,-10}
\begin{document}
		\textbf{\textit{Submitted to a journal for publication.}}

 \vspace{1cm}
{\centering
 
{\bfseries\Large Software-Defined Location Privacy Protection \\ for Vehicular Networks \bigskip}
 
Abdelwahab Boualouache, Ridha Soua, and Thomas Engel \\

\vspace{0.5cm}
   
   {\itshape
SnT, University of Luxembourg, Luxembourg\\
Email: \{abdelwahab.boualouache, ridha.soua, thomas.engel\}@uni.lu

   }
}

\begin{abstract}

While the adoption of connected vehicles is growing, security and privacy concerns are still the key barriers raised by society. These concerns mandate automakers and standardization groups to propose convenient solutions for privacy preservation. One of the main proposed solutions is the use of Pseudonym-Changing Strategies (PCSs). However, ETSI has recently published a technical report which highlights the absence of standardized and efficient PCSs \cite{etsi}. This alarming situation mandates an innovative shift in the way that the privacy of end-users is protected during their journey. Software Defined Networking (SDN) is emerging as a key 5G enabler to manage the network in a dynamic manner. SDN-enabled wireless networks are opening up new  programmable and highly-flexible privacy-aware solutions.
We exploit this paradigm to propose an innovative software-defined location privacy architecture for vehicular networks. The proposed architecture is context-aware, programmable, extensible, and able to encompass all existing and future pseudonym-changing strategies. To demonstrate the merit of our architecture, we consider a case study that involves four pseudonym-changing strategies, which we deploy over our architecture and compare with their static implementations. We also detail how the SDN controller dynamically switches between the strategies according to the context.





\end{abstract}


\textit{\textbf{Keywords: }}\noindent  5G, Vehicular Networks, Software Defined Networking, Location privacy, \\Pseudonym-Changing

\section{Introduction}

As part of the vision of 5G, connected vehicles will be an important pillar of Intelligent Transportation Systems (ITS), with the aim to ensure road safety, avoid traffic congestion and provide a better driving experience  for  users  during  their  journey. Although  the  deployment stage for connected  vehicles is imminent, many security and privacy issues are still unsolved. Location privacy is one of the main issues that may impede the wide acceptance of vehicular networks. Indeed, location tracking of  vehicles may reveal every place visited by drivers. This is because there is generally a one-to-one relationship between the vehicle and its driver.  The visited locations may include very personal places like hospitals, banks, insurance companies, etc. and hence can reveal sensitive information about the end-user. 

On the other hand, the main wireless communication technologies for connected vehicles, such as IEEE802.11p, present several privacy concerns. Indeed, IEEE802.11p mandates that each connected vehicle should frequently send a safety message, called CAM (Cooperative Awareness Message), to ensure cooperative awareness among neighboring vehicles. These messages include sensitive information such as identifiers, positions, speeds, etc, and are sent in clear text; hence vehicles could be tracked on the basis of the information transmitted by the Vehicle to Vehicle (V2V) and Vehicle to Infrastructure (V2I) communications.
To mitigate this privacy risk, the use of pseudonym schemes has received significant interest from the research community and standardization authorities. For instance, both the European standard ETSI TS 102 941 \cite{etsi} and the American standard SAE J2735 \cite{sae} have adopted a pseudonym scheme. However, several studies have shown that the use of a simple pseudonym-changing is insufficient to provide unlinkability between the pseudonyms and have suggested using strategies for changing the pseudonyms. Although, there is a  significant number of proposed Pseudonym-Changing Strategies (PCSs), there are no recommendations by standardization bodies for PCSs to apply. The main reasons for this can be summarized as follows: (i) several proposed strategies are  strongly topology-dependent i.e. they could only be applied in a given situation or area such as signalized intersections, parking lots, and gas stations; (ii) some strategies propose the use of radio silence without considering its critical impact on the exchange of safety messages, or dynamically readjusting radio silence duration; (iii) each PCS is evaluated with individual privacy metrics that may not be suitable for another PCS. This absence of unified evaluation metrics complicates the comparison between existing strategies; (iv) The scarcity of scientific studies focusing on the non-cooperation behavior of vehicles. 
Selfish vehicles could significantly decrease the efficiency of PCSs; (v) most of the existing PCSs assume a strong global passive adversary. However, this assumption is not realistic, since the global presence of the adversary is difficult to achieve due to the large scale of deployment of connected vehicles, and the high cost of ensuring complete coverage; and (vi) pseudonyms could be easily used to perform Sybil attacks. This vulnerability is not taken into the account by most of the strategies proposed in the literature.

One possible solution for dealing with these various PCSs is to propose a comprehensive architecture that is able to encompass all of them and their intrinsic features. This architecture should be forward-looking in the sense that it should support future PCS solutions. With this in mind, we propose a software-defined architecture for location privacy in vehicular networks. This architecture extends and leverages the concepts of SDN to PCSs and ensures the selection of the appropriate PCS according to the context of vehicles and other factors,  as will be detailed later. The SDN control plane orchestrates the selection and adjusts the  parameters of the selected strategy dynamically, based on information received from the data plane. The SDN strategy  rules are also forwarded from the control plane to the data plane to ensure the correct execution of the selected PCS. 
In the next section, we discuss in detail the standardization efforts related to pseudonym-changing strategies, their open issues, and explain how our proposed architecture cope with these issues.

\section{Pseudonym-Changing Strategies: Standardization Efforts and Open Issues}


Security standardization bodies have agreed to adopt PCS to protect the location privacy of connected vehicles. However, while in the US, the Society of Automotive Engineers (SAE) suggests that vehicles change their pseudonym every five minutes \cite{sae}, the European telecommunications standardization organization, ETSI, does not suggest the adoption of any PCS~\cite{etsi}. In the light of this, many PCSs are proposed in the literature. In \cite{mysurvey}, we presented a comprehensive survey and classification of these strategies. This paper also highlights open issues and presents recommendations, including the importance of developing a dynamic system to select the applying PCS according to the vehicular context.  Recently, ETSI published a technical report (ETSI TR 103 415) \cite{etsi} that presents a pre-standardization study of PCS. This document surveys the existing categories of strategies. It also discusses and describes the suggestions of the European projects (PRESERVE, SCOOP@F, and C2C-CC) regarding PCS. The document identifies the open issues of PCSs and proposes a set of recommendations addressing these issues. In the following, we discuss the open issues highlighted in~\cite{etsi,mysurvey} and the related recent advances

\begin{itemize}
    \item \textbf{Impact on road safety}: as shown in \cite{mysurvey}, strategies using radio silence are the most efficient solutions. However, their major drawback is their significant negative impact on safety-related applications. This was first investigated in \cite {safety1}, where the authors recommend that the silent period should be shorter than two seconds and that long silent periods can result in hazardous situations, since many safety messages will not be transmitted due to radio silence. 
The ETSI technical report \cite{etsi} also discusses the problems of ``missing vehicles" and ``guest vehicles". Missing vehicles are those that put radio silence into effect after changing their pseudonyms; at the end of this period, these vehicles suddenly appear in the LDMs (Local Dynamic Map) of neighboring vehicles. This may generate unpredictable reactions as highlighted in \cite{etsi}. In contrast, the problem of the guest vehicle is observed when a vehicle changes its pseudonym while his old pseudonym still populates the LDMs of its neighboring vehicles \cite{saftey2}. Subsequently, LDM messages contain two entries that correspond to the same vehicle, leading to a misinterpretation of the surrounding environment by neighboring vehicles. Unlike the missing vehicle problem, the ghost vehicle problem is not only linked to radio silence based strategies, but to PCSs in general.

\item \textbf{Non-cooperative behavior}: by triggering the change of their pseudonyms at the same time slot, cooperative vehicles ensure a high level of anonymity and create confusion for the attacker. Consequently, the existence of non-cooperative vehicles will significantly hinder the efficiency of the PCS specifically under lower vehicular density. The authors of \cite{noncooperative}  study PCSs under a non-cooperative environment. They propose a game theory model and find a Nash equilibrium of the PCS under different types of games (static/dynamic, with and without complete information). Other works such as \cite{motivationbased} and \cite{privanet} propose incentive mechanisms to motivate non-cooperative vehicles to participate in the PCS.

\item \textbf{Attacker model:} It is not trivial to estimate the power of tracking attackers that may exist in the future deployment of vehicular networks. Attacker power can be expressed in terms of tracking capabilities (strong or weak sniffing stations, efficiency of the tracking algorithm, etc.) and the coverage area. In addition, it is critical to properly define a realistic attacker model. For this reason, most of proposed PCSs have assumed the extreme case of the attacker model (global attacker full of capabilities); however, this assumption is not realistic, because global coverage entails a significant surveillance cost. Consequently, the authors of \cite{attacker_model} propose a mid-sized attacker whose power is in between that a local attacker and a global one. They also distinguish three tracking periods (i.e short-term, mid-term, and long-term) and two levels of surveillance granularity (i.e Road-level and Zone-level). 

\item \textbf{Evaluation metrics:} many metrics are proposed to assess the performance of PCSs. The recent study carried out by Zhao et al. \cite{privacy_metrics} show that there is no single privacy metric that outperforms all others under different contexts (mobility, traffic conditions, road section, etc.). 
For this reason, it is recommended to combine all metrics to obtain a fair performance evaluation of a PCS.

\item \textbf{Privacy model:} the privacy level depends mainly on the considered attacker model and the evaluation metrics. The authors of \cite{noncooperative} proposed a linear model to quantify the loss of privacy after the last change of pseudonym. In this model, the privacy level of vehicles linearly decreases according to a sensitivity parameter, which characterizes the power of the adversary. However, this model has two major drawbacks: (i) it does not specify how the sensitivity parameter is measured. (ii) the linearity of this model is not justified.

\item \textbf{Sybil attacks:} In this attack, vehicles use multiple identities, called Sybils, which can be exploited to create a fake traffic jam and hence to alter other vehicles' perceptions. Pseudonyms could be exploited to launch Sybil attacks. 
The ETSI technical report \cite{etsi} gives some recommendations on thwarting Sybil attacks, such as setting the maximum number of pseudonyms that can be used simultaneously and the minimum duration for which the pseudonyms should be used. The technical report also recommends the use of misbehavior detection systems.

\item \textbf{Pseudonym lock:} ETSI standards specify that the PCS could be locked on-demand for a maximum of $255$s, in particular when a critical safety situation occurs. The priority levels of such a situation are respectively ``0"  or ``1" \cite{etsi_saftey}. PCS locking is also proposed by the SAE. However, the conditions when the pseudonyms are locked are not yet defined.

\item \textbf{Pseudonym reuse:} Although the reuse of pseudonyms minimizes the storage capacity and facilitates the management of pseudonyms, it can  decrease the level of privacy. This is why the reuse of pseudonyms is not recommended as a privacy best practice. However, the Car2car consortium considers the reuse of pseudonym while defining some KPI to increase the privacy level \cite{mysurvey}.

\end{itemize}

It is obvious that several issues are still unsolved and need the attention of the research community to ensure the deployment of PCS in connected vehicles.
This urgent need motivates us to propose a forward-looking architecture that can encompass existing PCSs as well as future ones. Leveraging Software-Defined Networking (SDN), the proposed architecture is flexible and enables the efficient integration of new proposed solutions and new functions of the PCS. The contributions of this work can be  summarized as follows:

\begin{itemize}
    \item  Integration of SDN into vehicular networks, allowing new PCSs to be deployed, easily updated and dynamically reconfigured. 
    \item  Introduction of novel pseudonym-changing modules in the control plane and the definition of their different interactions to ensure a context-aware PCS.
    \item  Introduction of novel complementary pseudonym-changing modules in the data plane and the definition of their interactions.   
    \item  Definition of SDN rules (which can be modified dynamically at the controller) to establish a set of actions that will handle the PCS. 
    \item  Definition of a Sybil attack agent to interact with the external misbehavior system controller. 
    \item  Definition of self-learning module that is able to analyze and learn from its immediate context while autonomously adapting the PCS accordingly to ensure a high level of privacy protection. This module is crucial, as it guarantees network intelligence and leads to a network that is self-privacy-preserving.

\end{itemize}




\section{Proposed Architecture: Building blocks}

Our self-privacy-preserving architecture leverages the SDN paradigm and thus follows its main principle, which is the separation between the data and the control plane. The control plane is responsible for dynamically selecting the PCS, adjusting the parameters of strategy, and planning the strategy rules. On the other hand, the data plane translates the defined rules into actions to apply the PCS. The communications between the control plane and the data plane are secure.

\subsection{Control Plane}

\begin{figure*}[!ht]
\begin{center}
		\includegraphics[width=18cm,height=11cm]{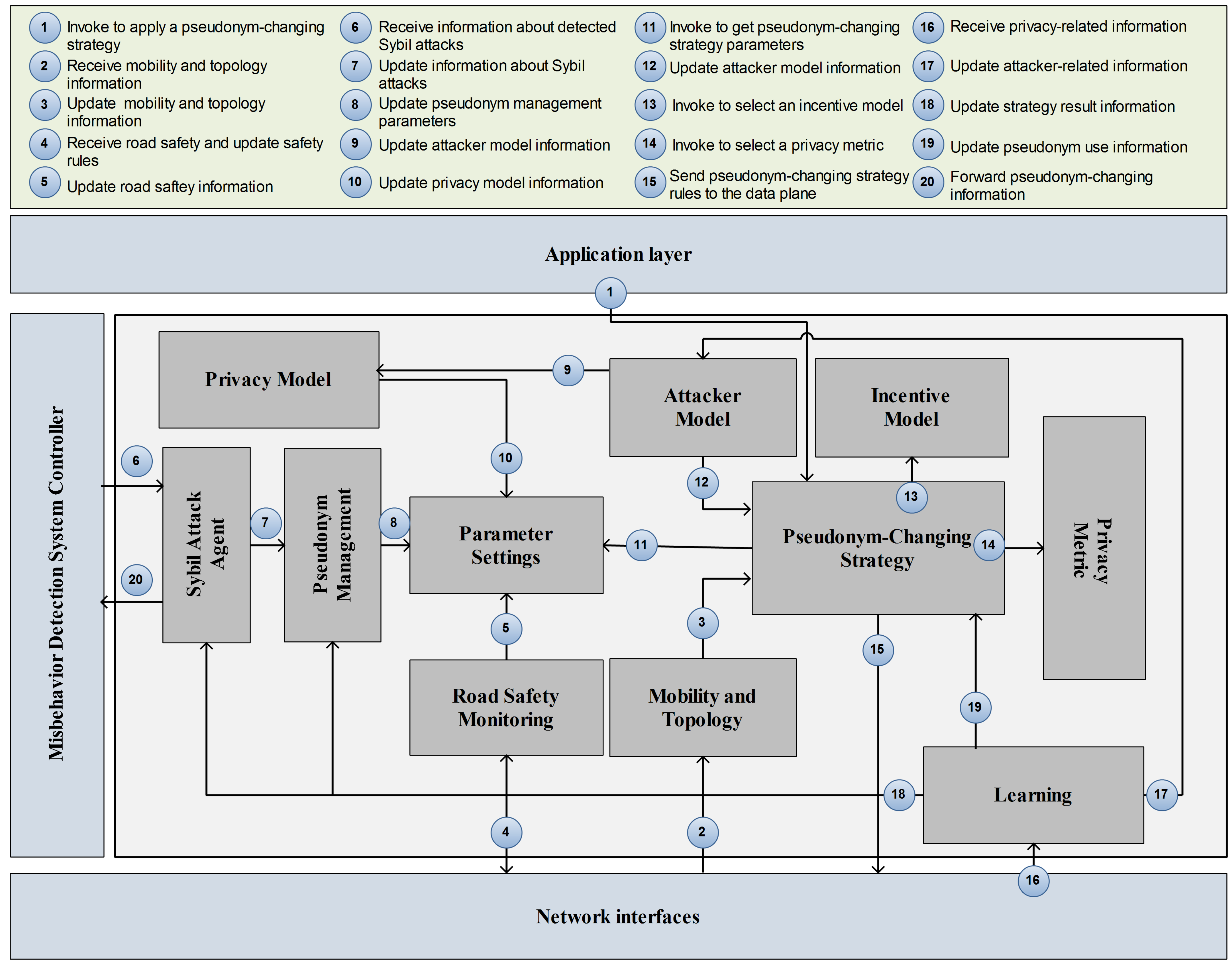}
	\end{center}
	\caption {The logical modules of the control plane and the interactions between them.}
	\label{fig:control_plane}
\end{figure*}

Figure~\ref{fig:control_plane} shows the logical modules of the control plane in our architecture. 
The PCS module receives a demand from the application layer to provide the location privacy service. This module chooses the most convenient PCS to be executed based on the information received from two modules: the Mobility and Topology module and Attacker Model module. Once the strategy is selected, the PCS module invokes (i) the Parameter Settings module to request the parameters of the strategy; (ii) the Incentive Model module to request the appropriate incentive method to motivate non-cooperative vehicles; and (iii) the Privacy Metric module to request indicators and KPIs for the evaluation of PCS performance. In the following, we detail these modules.

\begin{itemize}

	\item \textbf{Road Safety Monitoring:} this module monitors road conditions and its impact on traffic safety. Based on this assessment, the module develops appropriate SDN rules which are sent to the data plane. In addition, this module provides the necessary information to the Parameter Settings module to tune the PCS parameters, such as the duration of radio silence and the lock period. 
	
	\item \textbf{Misbehavior Detection System Controller:} this is an external component, which detects misbehaving attacks such as message injection, denial of service (DoS) and Sybil attacks. The SDN controller of our self-privacy-preserving architecture uses the information received from the Misbehavior Detection System Controller to update its parameters in order to limit Sybil attacks and returns information to help in detecting Sybil attacks and accurately evaluating the trust levels of vehicles.

	\item \textbf{Sybil Attack Agent:} this interface is used to interact with the Misbehavior Detection System Controller, receiving information from it and forwarding it to the Pseudonym Management module to adjust some PCS parameters. It also receives information from the Learning module and forwards this to the Misbehavior Detection System Controller to enhance the attack detection ratio.

	\item \textbf{Pseudonym Management:} this module plans the rules that  orchestrate the use of pseudonyms: the reuse of pseudonyms,  the frequency of changing of pseudonyms, the number of pseudonyms that can be used in parallel, etc. 
	This module receives information from both the Sybil Attack Agent and learning modules and sends the resulting rules to the Parameter Settings module. 
	
	\item \textbf{Privacy Model:} This is used to model the loss of privacy of vehicles over time. As explained in the previous section, the loss of privacy mainly depends on the strength of the attacker model. 
	For this reason, this module receives input from the Attacker Model module. The Privacy Model provides input to the Parameter Settings module, which in return specifies the parameters of the Privacy Model.

	\item \textbf{Mobility and Topology:} this module monitors the mobility pattern of vehicles and the road topology in real time. 
	
    \item \textbf{Parameter Settings:} this module sets the different parameters of the PCS, such as the duration of the radio silence period and the minimum duration of the use of pseudonyms. The definition of these parameters is made according to the information received from the Road Safety, Pseudonym Management, and the Privacy Model modules. 
    
    \item \textbf{Attacker Model:} this module evaluates the power of the attacker. As discussed in the previous section, the attacker can be internal or external, local or mid-sized, long-term. 
    It can perform simple syntactic linking of pseudonyms, but can also carry out more advanced semantic linking of pseudonyms. 
    This module gets regular updates from the learning model and sends feedback to the Pseudonym-Changing Strategy module.  
    
    \item \textbf{Incentive Model:} this module defines the incentive model, which is used to motivate selfish vehicles to participate in the PCS.
    
    \item \textbf{Privacy Metric:} this module defines the privacy metrics used to evaluate the PCS. It worth mentioning that the privacy metrics can be selected by the PCS to evaluate its own performance.
    
    \item \textbf{PCS Module:} this module defines the strategy to be executed based on the information received from the Mobility and Topology module and the Attacker Model module. Once the strategy is selected, this module invokes the Parameter Settings module to obtain the most appropriate parameters of the selected strategy. This module also invokes the Privacy Metric module and the Incentive Model module to select the evaluation metric and the incentive method respectively. 
    
   \item \textbf{Learning:} this module periodically receives privacy-related information from the data plane (i.e the privacy levels of vehicles, the presence of an attacker, and the set of selfish vehicles). This information is analyzed and forwarded to the corresponding modules: (i) the Attacker Model module to adjust the attacker model being used; (ii) the PCS module to tune the strategy parameters, and the Incentive Model module, and to select an additional potential privacy metric.
   (ii) the Pseudonym Management module to adjust pseudonym management related parameters, and finally (iv) the Sybil Attack Agent, which forwards pseudonym-changing information to the Misbehavior Detection System Controller. The purpose is to support this controller in the accurate detection of Sybil attacks and trust assessment of vehicles.

\end{itemize}

	\begin{figure*}[!h]
\begin{center}
		\includegraphics[width=18cm,height=10cm]{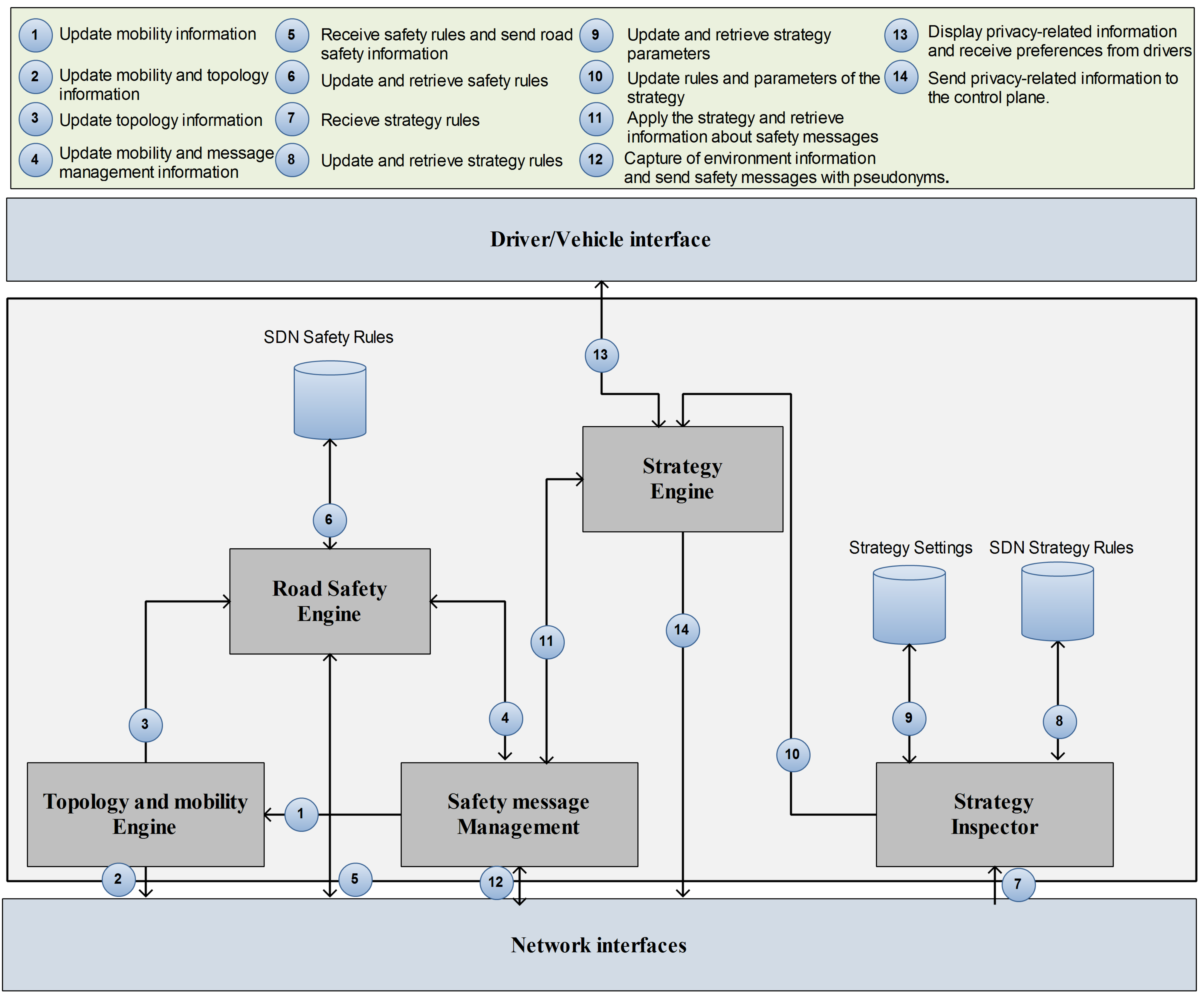}
	\end{center}
	\caption {The logical components of the data plane and the interactions between them.}
	\label{fig:data_plane}
\end{figure*}

\subsection{Data Plane}
The data plane is composed of the different vehicles that are involved in the PCS.
Figure~\ref{fig:data_plane} depicts the modules of the data plane, which are responsible of the execution of the PCS. The data plane uses the vehicles' communication interfaces to collect pertinent information concerns the surrounding vehicular environment. The data plane sends mobility, safety, and privacy information to the control plane, while it receives safety and strategy rules. In the following, we describe the modules and the databases of the data plane: 

\begin{itemize}

	\item \textbf{Safety Message Management:} this module sends and receives pseudonymous safety messages. It also receives instructions from the Strategy Engine. These instructions vary according to the applied strategy. In addition, this module provides the status of the surrounding environment and the impact of the applied PCS to the Road Safety Engine, the Topology and the Mobility engine, and finally to the Strategy Engine.
     
	\item \textbf{Mobility and Topology Engine:} Equipped with a map and GPS, this module sends the mobility information of the vehicle such as position, speed, and acceleration and the topology information to the Road Safety Engine and to the control plane. 
	
	
     \item \textbf{SDN Safety Rules:} This is a database, which contains the safety rules that are used to assess road conditions. The rules data is received from the control plane. 
     
     
	\item \textbf{Road Safety Engine:} this module receives, stores and updates the safety rules received from the control plane. These rules are used to evaluate road safety based on the information received from the Topology and Mobility Engine and the Safety Message Management module. This module periodically sends road safety information to the control plane.
	
	
	\item \textbf{SDN Strategy Rules:} This is a database that contains the rules related to PCS. These rules describe where, when and how pseudonyms change. The database is regularly updated by the Strategy Inspector module; based on the information received from the control plane.
	
    \item \textbf{Strategy Settings:} This is a database that contains the settings of the applied strategy such as the duration of radio silence period after the changing of pseudonym. This database is also regularly updated by the Strategy  Inspector module according to the information received from the control plane.
	
	\item \textbf{Strategy Inspector:} this module represents an interface, which communicates with the PCS module of the control plane. It receives information from the SDN controller(s) and stores them in two databases: the SDN Strategy rules and the Strategy Settings databases. This module also forwards these PCS rules and settings to the Strategy Engine module.
	
	\item \textbf{Strategy Engine:} this module executes the PCS according to the rules and settings received from the Strategy Inspector module. To execute the strategy, the module continuously monitors and sends instructions to the Safety Message Management module. This module provides privacy protection related information to the driver from whom it receives privacy level recommendations. This module also sends privacy-related information to the control plane. 
	
	\end{itemize}


\section{Case Study}
To demonstrate the merit of our proposed architecture, we conducted the following case study. As shown in Figure~\ref{fig:select_pcs} (1), we populated a Software-Defined Location Privacy Controller (SDLP) with four state-of-the-art PCSs: UPCS \cite{upcs}, TAPCS \cite{tapcs}, PRIVANET \cite{privanet} and SocialSpots \cite{social}. In this section, we first show how these strategies are integrated into our architecture. Then, we illustrate how the SDLP performs a context-aware PCS selection. The context is mainly defined by mobility and topology, as well as the attacker model. Finally, we conduct a simulation-based study to demonstrate how our proposed architecture dynamically updates the security parameters of each strategy.

\subsection{PCSs Deployment}


\begin{table*}[!h]
\caption {The deployments of PCSs in the self-privacy-preserving architecture}
\label{tab:pcs_deployement}
\resizebox{\textwidth}{!}{\begin{tabular}{|l|l|l|l|l|l|c|l|c|}
\hline
\rowcolor[HTML]{EFEFEF} 
\multicolumn{2}{|l|}{\cellcolor[HTML]{EFEFEF}}                                                                     & \multicolumn{2}{l|}{\cellcolor[HTML]{EFEFEF}\textbf{Mobility and topology}}                                                                    & \textbf{Parameter setting}                                                                                    & \textbf{Attacker model}                                                                                                                                                                          & \textbf{Privacy model} & \textbf{Privacy metric}                         & \textbf{Incentive model} \\ \hline
\multicolumn{2}{|l|}{\cellcolor[HTML]{EFEFEF}\textbf{UPCS \cite{upcs}}}                           & \multicolumn{2}{l|}{Signalized intersection}                                                                           & \begin{tabular}[c]{@{}l@{}}Red traffic light duration:\\ 30s, 60s\end{tabular}                                & {\color[HTML]{000000} \begin{tabular}[c]{@{}l@{}}Global external passive\\ and local internal passive\\ (Semantic and syntactic linking)\end{tabular}}                   & No                     & The entropy of the annonymity set               & No                       \\ \hline
\multicolumn{2}{|l|}{\cellcolor[HTML]{EFEFEF}\textbf{SocialSpots \cite{social}}}                  & \multicolumn{2}{l|}{Signalized intersection}                                                                           & Red traffic light turns green                                                                                 & {\color[HTML]{000000} \begin{tabular}[c]{@{}l@{}}Global external passive\\ (Syntactic linking)\end{tabular}}                                                             & No                     & The size of the anonymity set                   & Yes                      \\ \hline
\multicolumn{2}{|l|}{\cellcolor[HTML]{EFEFEF}\textbf{TAPCS \cite{tapcs}}}                         & \multicolumn{2}{l|}{Traffic congestion}                                                                                & Speed threshold                                                                                             & \cellcolor[HTML]{FFFFFF}{\color[HTML]{000000} \begin{tabular}[c]{@{}l@{}}Global external passive\\ and local internal passive\\ (Semantic and syntactic linking)\end{tabular}}                   & No                     & The entropy of the anonymity set               & No                       \\ \hline
\multicolumn{2}{|l|}{\cellcolor[HTML]{EFEFEF}}                                                                     & \multicolumn{2}{l|}{}                                                                                                  &                                                                                                               & \cellcolor[HTML]{FFFFFF}{\color[HTML]{000000} }                                                                                                                                                  &                        &                                                 &                          \\
\multicolumn{2}{|l|}{\cellcolor[HTML]{EFEFEF}\textbf{PRIVANET \cite{privanet}}} & \multicolumn{2}{l|}{\multirow{-2}{*}{\begin{tabular}[c]{@{}l@{}}Roadside Infrastructure\\ e.g. Gas station\end{tabular}}} & \begin{tabular}[c]{@{}l@{}} The capacity of RI \\ The threshold of privacy\end{tabular} & \cellcolor[HTML]{FFFFFF}{\color[HTML]{000000} \begin{tabular}[c]{@{}l@{}}Global external passive\\ and local internal passive\\ (Semantic and syntactic linking)\end{tabular}} & \multirow{-2}{*}{Yes}  & \multirow{-2}{*}{The size of the anonymity set} & \multirow{-2}{*}{Yes}    \\ \hline
\end{tabular}}
\end{table*}

\begin{figure*}[!h]
\begin{center}
		\includegraphics[width=18cm,height=10cm]{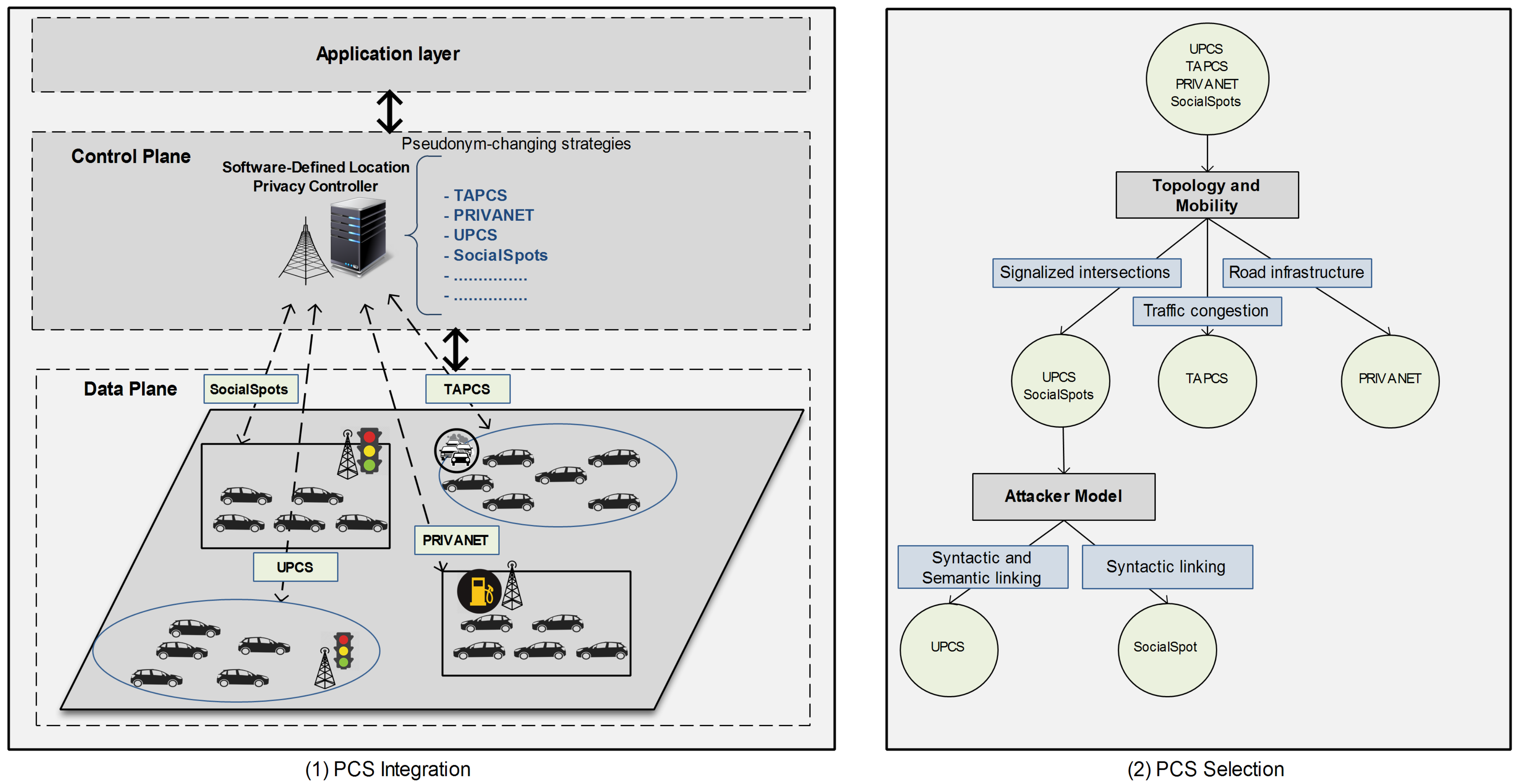}
	\end{center}
	\caption {The selection of pseudonym changing strategy}
	\label{fig:select_pcs}
\end{figure*}

Our proposed architecture is flexible enough to support any state-of-the-art PCS. Table~\ref{tab:pcs_deployement} shows how the considered strategies are mapped to our architecture. This table has six columns: (i) Mobility and topology: specifies the topology where the strategy can be used; (ii) Parameter Setting: specifies the parameters of the strategy; (iii) Attacker model: specifies that attacker model from which the strategy provides protection; (iv) Privacy model: specifies if the strategy uses a privacy model or not; (v) Privacy metric: specifies the metric used to evaluate the strategy; (vi) Incentive model: specifies if the strategy uses an incentive model or not.

Control plane modules are activated or deactivated according to the requirements of each PCS. For example, the Incentive Model module is disabled for UPCS and TAPCS since these strategies do not propose any mechanism to motivate non-cooperative vehicles to change their pseudonyms; while the Privacy Model module is only activated for PRIVANET strategy.


Figure~\ref{fig:select_pcs} (2) illustrates different steps for the selection of a PCS. The SDLP first checks information received from the Mobility and Topology module. For instance, if the vehicle is entering a signalized intersection, two PCSs could be applied to this context: UPCS and SocialSpots. To decide which of the two strategies to apply, SDLP checks information received from the Attacker Model module. If the attacker model can perform both syntactic and semantic pseudonym linking attacks, then UPCS is selected. Otherwise, if the attacker can perform only syntactic attacks, SocialSpots is selected. More details on syntactic and semantic pseudonym linking attacks can be found in \cite{mysurvey}.

\begin{table*}[]
\caption {The configuration of pseudonym-changing strategies}
\label{tab:pcs_configuration}
\begin{tabular}{|l|l|l|l|l|}
\hline
\rowcolor[HTML]{EFEFEF} 
\multicolumn{1}{|c|}{\cellcolor[HTML]{EFEFEF}\textbf{Strategy}} & \textbf{Changed context} & \textbf{Configuration} & \textbf{Action} & \textbf{Results} \\ \hline
\cellcolor[HTML]{EFEFEF} &  & \begin{tabular}[c]{@{}l@{}}10\% of vehicles \\ in dangerous situation\end{tabular} & Pseudonym lock & \begin{tabular}[c]{@{}l@{}}Low safety risk\\ Acceptable privacy level\end{tabular} \\ \cline{3-5} 
\multirow{-2}{*}{\cellcolor[HTML]{EFEFEF}\textbf{SDN-based UPCS \cite{upcs}}} & \multirow{-2}{*}{Road safety} & \begin{tabular}[c]{@{}l@{}}20\% of vehicles \\ are in dangerous situation\end{tabular} & Pseudonym lock & \begin{tabular}[c]{@{}l@{}}Low safety risk.\\ Acceptable privacy level\end{tabular} \\ \hline
\cellcolor[HTML]{EFEFEF} &  & Simple attacker & Select privacy metric & The size of the anonymity set \\ \cline{3-5} 
\cellcolor[HTML]{EFEFEF} &  & Medium attacker & Change the privacy metric & The entropy of the anonymity set \\ \cline{3-5} 
\multirow{-3}{*}{\cellcolor[HTML]{EFEFEF}\textbf{SDN-based TAPCS \cite{tapcs}}} & \multirow{-3}{*}{Attacker model} & Advanced attacker & Keep the privacy metric & The entropy of the anonymity set \\ \hline
\cellcolor[HTML]{EFEFEF} &  & Sensitivity parameter = 0.1 & Update privacy model & High privacy level \\ \cline{3-5} 
\multirow{-2}{*}{\cellcolor[HTML]{EFEFEF}\textbf{SDN-based PRIVANET \cite{privanet}}} & \multirow{-2}{*}{Privacy model} & Sensitivity parameter = 0.2 & Update privacy model & Low privacy level \\ \hline
\end{tabular}
\end{table*}

\subsection{Simulation Setup}
 We carried out a simulation-based analysis to demonstrate the merit of our SDN-based and self-learning architecture and how it dynamically adapts PCS security parameters to the context. This simulation-based analysis was performed using Veins Simulation Framework \cite{veins}. The considered scenario is similar to that proposed in \cite{privanet}. Three strategies are simulated: UPCS, TAPCS, and PRIVANET. SocialSpots was excluded, as it has the same application context (signalized intersections) as UPCS.
 
Table~\ref{tab:pcs_configuration} details the configurations of the simulated strategies. This table has four columns: (i) Changed context: specifies the context we change during the simulation; (ii) Configuration: specifies the values we assign to the context' parameters; (iii) Action: specifies the action to perform when the parameter is changed; (iv) Results: specifies the obtained results when the action is applied. To demonstrate the dynamic changing of PCS parameters according to context, three different scenarios are considered. 
\begin{enumerate}
    \item \textbf{Scenario 1}: uses UPCS strategy in a road safety context, where the number of vehicles in a dangerous situation can be 10\% or 20\%.  The pseudonym changing in such a situation can generate traffic collisions and accidents.
    \item \textbf{Scenario 2}: uses TAPCS strategy, where we study how this strategy adapts the privacy metric to the attacker model. Three configurations of the attacker model are considered: simple, medium, and advanced. 
    \item \textbf{Scenario 3}: uses PRIVANET focusing on the privacy model. We consider two configurations of this model by varying the sensitivity parameter value, which characterizes the power of the adversary.
    
\end{enumerate}

\subsection{Simulation Results}

\begin{figure}[!h]
\begin{center}
		\includegraphics[width=9cm,height=8cm]{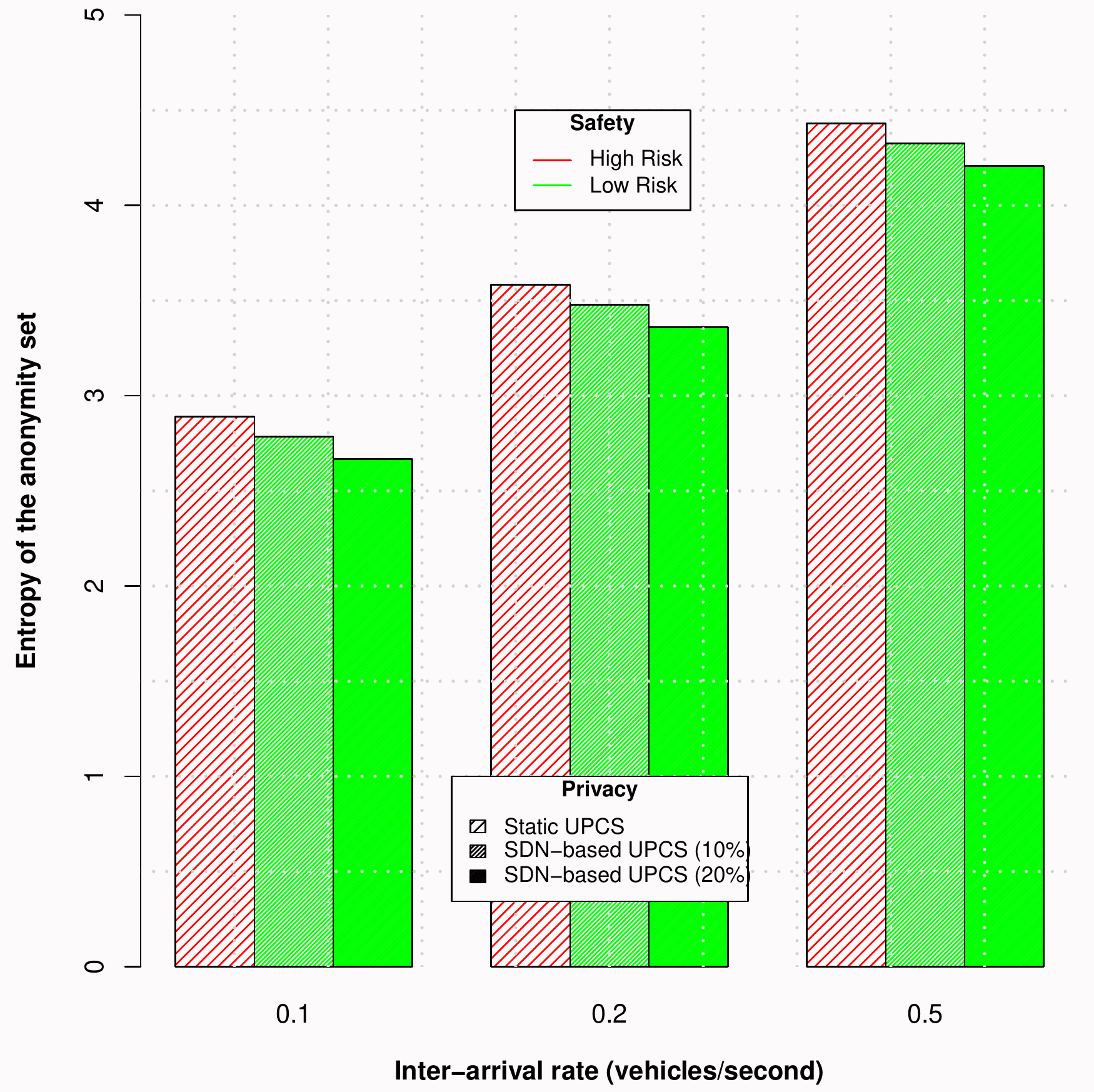}
	\end{center}
	\caption {Static UPCS vs SDN-based UPCS}
	\label{fig:upcs}
\end{figure}

Figure~\ref{fig:upcs} compares the static implementation UPCS (static UPCS) to its SDN-based variant (SDN-based UPCS). Two performance indicators are considered: the privacy level and safety. 
As shown in Figure~\ref{fig:upcs}, static UPCS provides a higher level of privacy protection compared to SDN-based UPCS. However, SDN-based UPCS has a lower safety risk than static UPCS. The reason for this, as described in Table~\ref{tab:pcs_configuration}, is that SDLP takes an action to lock pseudonym-changing processes of vehicles in a dangerous situation. This lock slightly decreases the privacy protection level, while reducing the safety risk.

\begin{figure}[!h]
\begin{center}
		\includegraphics[width=9cm,height=8cm]{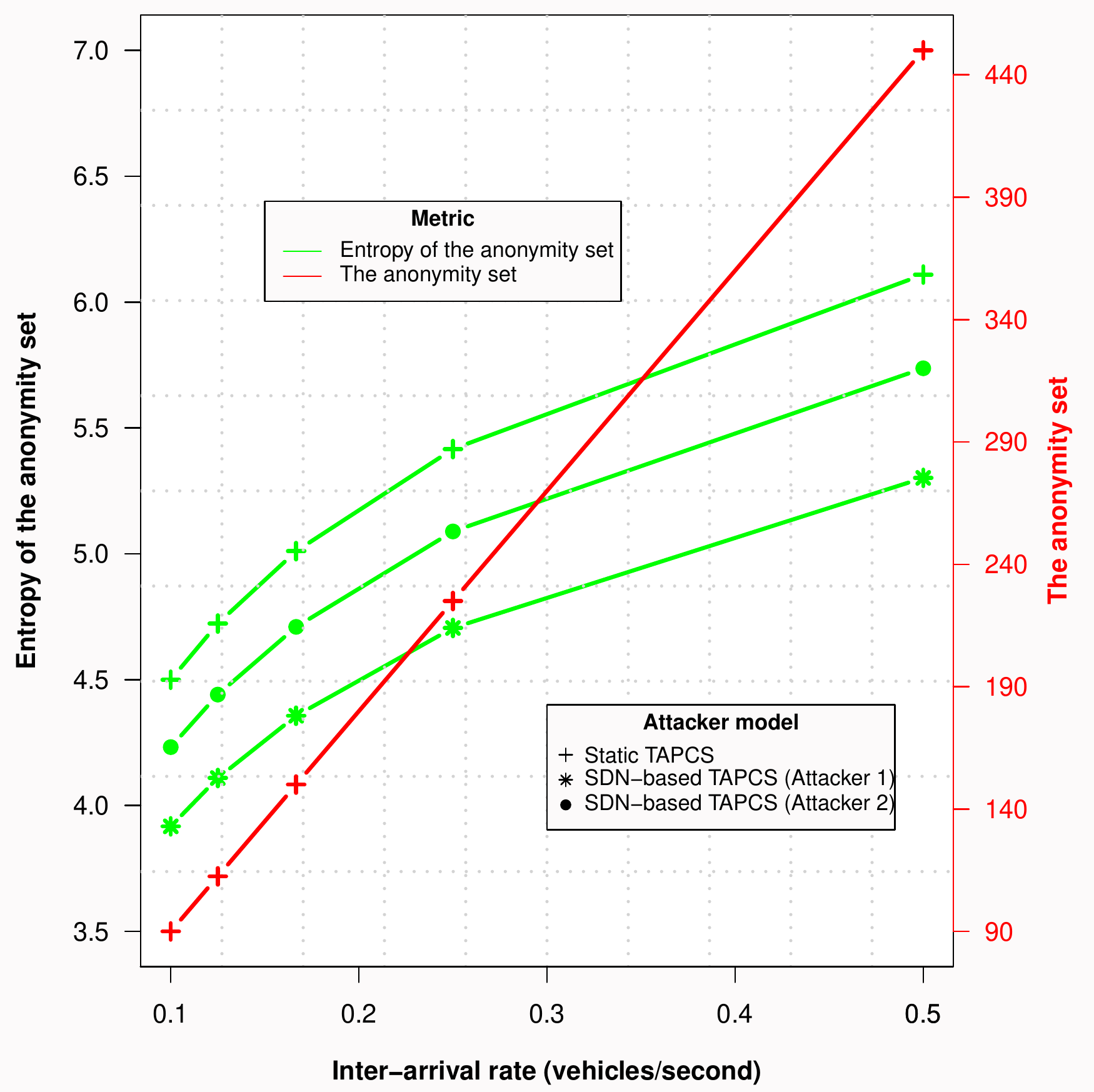}
	\end{center}
	\caption {Static TAPCS vs SDN-based TAPCS}
	\label{fig:tapcs}
\end{figure}

Figure~\ref{fig:tapcs} makes a comparison between Static TAPCS and SDN-based TAPCS. In Static TAPCS, the entropy of anonymity set is used as a performance metric, whatever attacker model is used. However, the SDN-based TAPCS varies the performance metric according to the power of the attacker. For instance, the size of the anonymity set is chosen when the attacker is simple, while the entropy of the anonymity set is selected when the attacker is medium or advanced. In the former case, the probabilities of distinguishing between vehicles are equal and hence the measuring size of the anonymity set performs well. In the latter case, the distinguishing probabilities are not equal; hence the need to take the entropy into account.

\begin{figure}[!h]
\begin{center}
		\includegraphics[width=9cm,height=8cm]{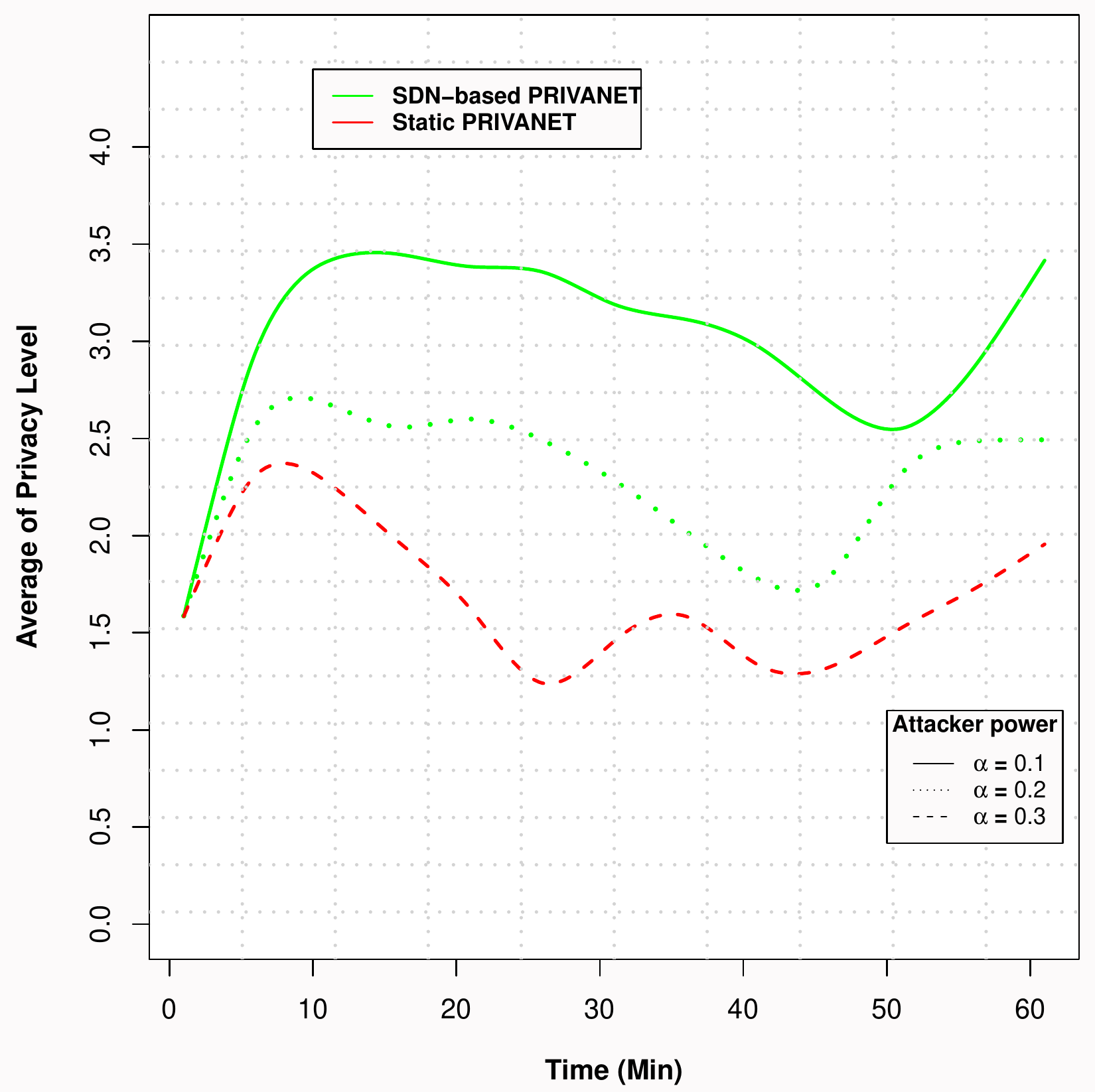}
	\end{center}
	\caption {Static PRIVANET vs. SDN-based PRIVANET}
	\label{fig:privanet}
\end{figure}

Finally, we compare the static implementation of PRIVANET and the SDN-based PRIVANET. As illustrated in Figure~\ref{fig:privanet}, the sensitivity parameter ($\alpha$), which characterizes the power of the attacker, remains unchanged in Static PRIVANET  and is equal to $0.3$. However, for SDN-based PRIVANET, the sensitivity parameter is updated according to the information received from the data plane. The change in the power of the attacker (the sensitivity parameter) has a direct impact on the privacy level obtained by vehicles. Indeed, as illustrated in  Figure~\ref{fig:privanet}, the high values of the average of privacy are obtained when the sensitivity parameter equals 0.1. However, the lower values of the average of privacy are obtained when the sensitivity parameter equals 0.3.


\section{Conclusion}
The imminent deployment of connected vehicles requires significant attention to the security and privacy aspects. Privacy protection is a critical issue that influences the user acceptance of this technology. Pseudonym-changing strategies are considered as the key solution to overcome this acute need. However, the absence of recommended pseudonym-changing strategies (PCSs) represents an obstacle to achieving this objective. To this end, we propose an innovative architecture that exploits Software-Defined Networking (SDN), one of the key technologies for 5G networks. Our proposed architecture is flexible and self-learning and hence can encompass PCSs proposed so far in the literature and even upcoming PCSs. The selection of the appropriate PCS and it security settings are context-aware. The control plane is modular and includes the main building-blocks of PCSs which can support any future solution.  


\section*{Acknowledgment}
This work was supported by the H2020 5G-DRIVE project (ID: 814956) and the H2020 5G-MOBIX project (ID: 825496).

\bibliographystyle{IEEEtran}
\bibliography{IEEEabrv,main}

\end{document}